\documentclass[epj]{svjour}
\usepackage{amssymb,amsmath}
\usepackage{graphicx,float}
\begin{document}

\title{ Jamming  and Correlation Patterns in Traffic of Information on Sparse Modular Networks}
\author{Bosiljka Tadi\'c\inst{1} \and Marija Mitrovi\'c\inst{2} }%
%
\institute{Department  for Theoretical Physics, Jo\v{z}ef Stefan Institute, 
P.O. Box 3000, SI-1001 Ljubljana, Slovenia,  \and Scientific Computing Laboratory, Institute of Physics,  Belgrade, Serbia}
\date{Received: date / Revised version: date}

\abstract{
We study high-density traffic of information packets on sparse modular networks with scale-free subgraphs. With different statistical measures we distinguish between the free flow and congested regime and point out the role of modules in the jamming transition. We further consider correlations between traffic signals collected at each node in the network. The correlation matrix between pairs of signals reflects the network modularity in the eigenvalue spectrum and the structure of eigenvectors. The internal structure of the modules  has an important role in the diffusion dynamics, leading to enhanced correlations between the modular hubs, which can not be filtered out by standard methods. Implications for the analysis of real networks with unknown modular structure are discussed.
 \PACS{
      {89.75.Hc}{Networks and genealogical trees}   \and
      {05.40.Ca}{Noise}\and
      {02.70.-c}{Computational techniques; simulations}
     } 
}
\titlerunning{Traffic Jamming \& Correlations Patterns}
\maketitle
\section{Introduction}
In recent years network research has been intensified aiming at quantitative 
representation and study  of the interactions in  complex dynamical systems \cite{boccaletti2006}.  
These networks often exhibit hidden structures and inhomogeneity at mesoscopic scales. Subgraphs of different sizes and topological consistency often appear in real networks, such as  modules or motifs in gene networks \cite{motifs}, community structure in social networks \cite{danon2006}, topological clusters  or dynamical aggregation on the Internet \cite{flake}, and others. It has been recognized  that in the evolving networks  functional units have emerged,  and that in different functional networks they may be represented by topologically characteristic subgraphs, e.g.,  communities, modules, paths, trees, etc. 
Subgraphs on modular networks can be  recognized topologically by better or tighter connected group of nodes \cite{danon2006}. Sparseness of real networks is another feature which is tightly connected with the network dynamic stability: Large connectivity may induce a chaotic behavior (positive Lyapunov exponent) in networks even for simple dynamics of its unites \cite{dejan_sparse}. This might be a part of the reason why most of the networks in nature (except perhaps brain) are sparsely connected \cite{boccaletti2006,gardner2005,yeung2002,Costa2008}. 
For these reasons it is of great importance  to understand the inter-relationships between dynamics and structure in {\it sparse modular  networks} \cite{mmbt2008}. This question is also in the focus of the present work. We study transport processes on  networks with sparsely connected modules by means of the numerical simulations of dense traffic, and spectral analysis of the Laplacian matrix and matrices generated by the correlations between the traffic time series.  

{\it Spectral analysis.} Important information about complex network structure and dynamical processes is contained in the  eigenvalue spectra  and corresponding eigenvectors of the adjacency matrix and of other, e.g., Laplacian matrices related to its structure\cite{boccaletti2006,samukhin2007,farkas2001,mmbt2008}. Recent studies of the synchronization of phase-coupled oscillators \cite{boccaletti2006,danon2006,mcgraw2008,jurgen2005} in modular networks have revealed strict relationship between the  synchronization and smallest nonzero eigenvalues of Laplacian matrix \cite{arenas_guliera2006}. Furthermore, positive/negative components of the corresponding eigenvectors appear to be well localized on the modules \cite{boccaletti2006,donetti2004}. Other types of diffusive dynamics,  like spreading of disease \cite{eigen_cen} and traffic or navigated  walks \cite{tadic2007,bt-arw01,guimera2002,nr-rwcn-04} are often studied on different networks. The exact relationship between the return-times distribution of the random walk on networks and the spectral properties of the respective Laplacian matrix, has been derived theoretically \cite{samukhin2007} and confirmed numerically \cite{mmbt2008} for the tree graphs. Furthermore, in \cite{mmbt2008} the Laplacian spectra have been studied in detail for  a wide class of sparse and modular networks.

{\it Time series correlations.} In studies of complex networks much effort has been  invested in understanding how the structure of a network is manifested in network function. Beside its theoretical meaning, this question has a great practical importance. For instance, in bio-engineering \cite{collins2003} and neurosciences \cite{jurgen},  one faces the problem to design a network with given function, or to reconstruct a network from its measured dynamical output. Often the empirical data are available as  {\it time series collected at different nodes} and within specified time windows, for instance in the market dynamics \cite{mantegna1999} or gene expressions \cite{zivkovic2006}, and other. Different methods have been employed, for instance, the correlation matrix reconstruction of neural or anti-gene interaction network, \cite{eshel2004,eshel2008},  gene network reconstruction algorithms based on SVD and assumed gene dynamics model \cite{dejan_metod,collins2003}. 
The goal of the reversed engineering is to unravel interactions  between the nodes  which are the cause of the observed dynamical output (time series). Apart from often limited information, this is a hard problem to which both the network structural complexity and nonlinearity of the dynamics contribute.  The presence of the  modules and other mesoscopic inhomogeneities representing functional  groups of nodes, may increase the complexity of the problem.  
In principle, the inherent consistency of these approaches may rely on certain robustness of the inter-dependences between the network structural elements and  the properties of the time series,  seen within given type of the dynamics. 
In particular, time series from the diffusion processes, e.g. synchronization in neural networks, are considerably different  from the autocatalytic regulation in gene networks. The diversity in the activity of nodes in the case of random-walk dynamics on structured networks is directly related to the node connectivity.
One of the goals in this paper is to examine the efficiency and limitations of the network reconstruction form the traffic time series  on modular network. For this purpose we run known dynamics (traffic of information packets with queuing \cite{tadic2007}) on known network structure and record the traffic time series at all nodes. Then we construct the correlation matrix of these time series and use the standard filtering methods to uncover the structure behind the correlations. The degree of similarity  between the filtered correlation matrix and the original adjacency matrix,  as well as between their eigenvalue spectra is quantified.

The paper is organize as follows: The modular network structures are introduced and results of simulations of traffic of information packets on these  networks are presented in Section\ \ref{sec:traffic}.  Traffic properties near the jamming are studied in detail by statistical means. In Section\ \ref{sec:correl} we present the construction and filtering of the correlation matrix from the traffic time series. Furthermore, a detailed  spectral analysis of the Laplacian matrices of the original networks and the correlation networks is given in Section\ \ref{sec:correl-spectra}.
 A short  summary and the discussion of the results are presented  Section\ \ref{sec:conclusion}.

\section{Traffic Jamming on Modular Networks} \label{sec:traffic}

\subsection{Network Structures}
For simulations of the information traffic we use two types of modular networks shown in Fig.\ \ref{networks}, in particular: (a) the network composed of few large modules with random connections between them (Net269), and (b) network consisting of a large number of smaller modules linked through a scale-free tree graph. These networks are grown using the algorithms which are introduced in \cite{mmbt2008}.
 The structural properties of these networks are controlled by  three  parameters: the average connectivity $M$, the probability of new module $P_0$, and the attractivity of node $\alpha$ which controls rewiring process during the module growth. By the numerical implementation and choosing the values of these parameters the internal structure of groups (modules) as well as the structure of the network connecting different modules can be varied in a desired manner (see Ref\cite{mmbt2008} for details). Specifically,  for the networks shown in Fig.\ \ref{networks} have $N=1000$ nodes and the  values of the control parameters are as follows: 
\begin{figure}[htb]
\begin{center}
\begin{tabular}{c} 
\resizebox{18pc}{!}{\includegraphics{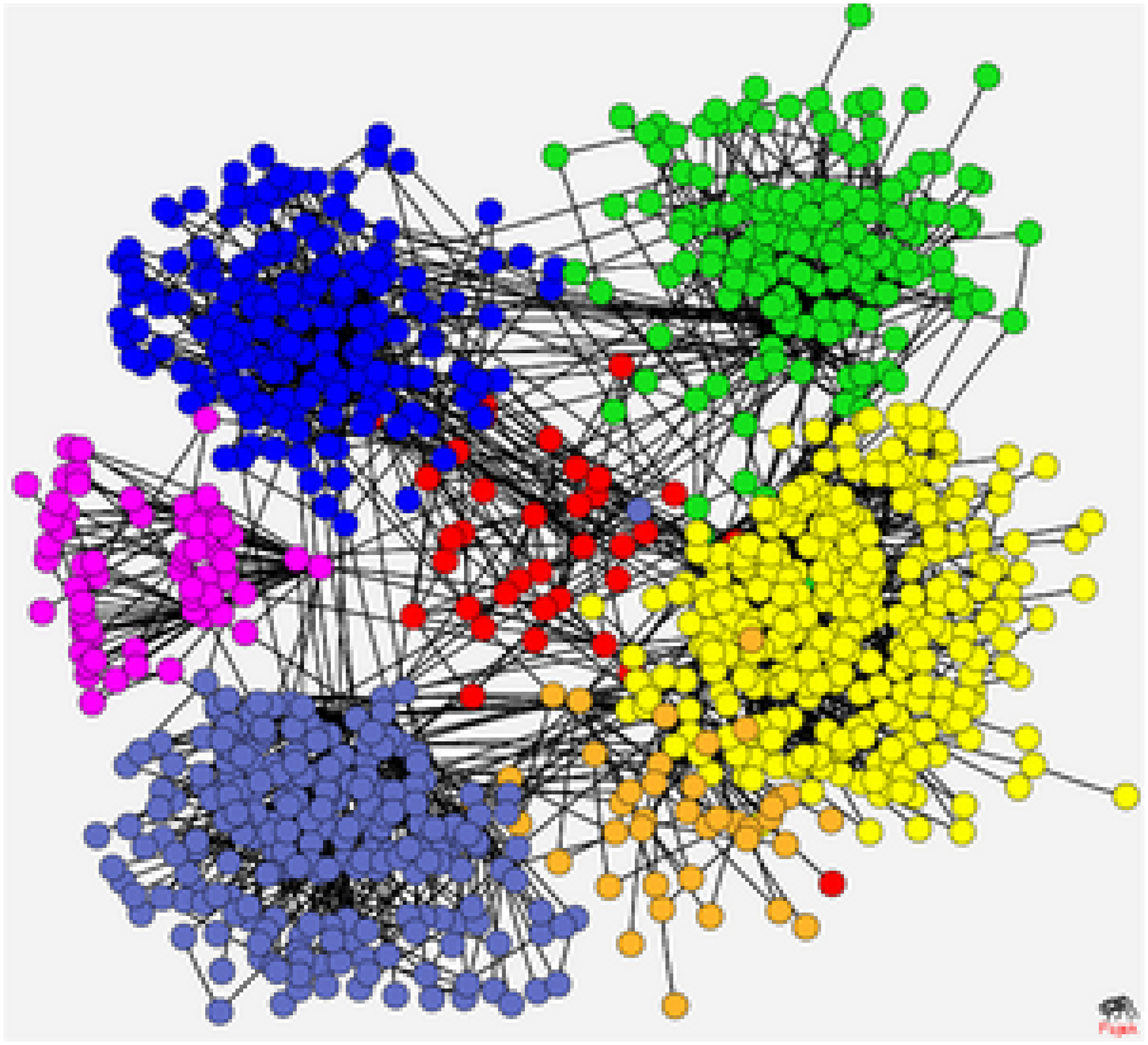}}\\
\resizebox{18pc}{!}{\includegraphics{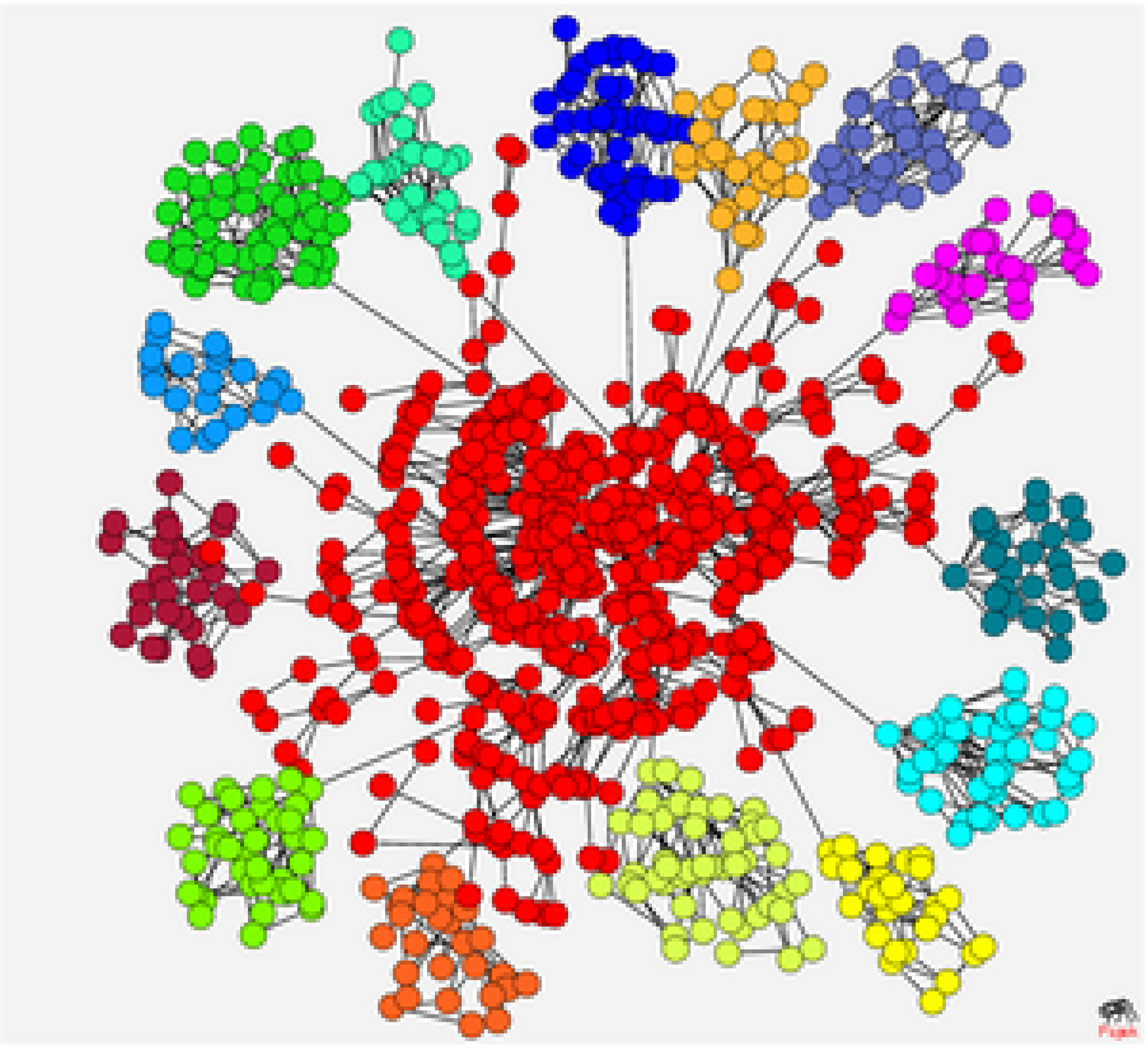}}\\
\end{tabular}
\end{center}
\caption {(top) Original network with large interconnected scale-free modules,  Net269, and (bottom)  Scale-free tree structure with attached  scale-free modules. Colors indicate membership of nodes to topologically distinct subgraphs.}
\label{networks}
\end{figure} 
$M=2$, $10\%$ of rewired links ($\alpha=0.9$) and $P_0=0.006$ for Fig.\ \ref{networks}(top), referred as  Net269. In different structure, shown in Fig.\ \ref{networks}(bottom), we combine scale free tree ($M=1$, $\alpha =1$)  with a number of smaller modules, where each module is grown taking the same parameters  $M=2$, $\alpha =0.9$. The modules are attached to  ending nodes of the scale free tree Fig.\ \ref{networks}(bottom). Note that the internal structure of these modules is statistically the same as in case of Net269, however their size  varies between $20$ and $50$ nodes and their number is chosen in such a way that approximately half of total number of nodes are members of the modules, while the other half of nodes belong to the underlying tree structure. It should be stressed that, in contrast to the Net269 in Fig.\ \ref{networks}(top), which can be fully partitioned into modules (communities), the modular network in Fig.\ \ref{networks}(bottom) consists of modules and the underlying network (tree), which is structured itself. In the following sections we will show how these two components of the modular networks affect the traffic and study the patterns of correlations in the traffic signals.

\subsection{Traffic of Information Packets}

In this Section we simulate dense traffic of information packets on the network structures introduced above. We use our traffic model introduced earlier (see review article  \cite{tadic2007} for details). We describe in short the main features of the traffic model and the relevant parameters (details of the numerical implementation of the rules are given in \cite{bt-lncs2003}):
\begin{itemize}
\item {\it Creation and assignment.} At each time step each node creates a packet with a given rate $R$ and assignees it a randomly selected recipient node (delivery address). 

\item {\it Navigation.} Each node processes a packet from top of its queue (LIFO-queue) 
towards one of its neighbours. The neighbour node is selected according to {\it nnn-}navigation rule \cite{tadic2005,tadic2007}, in which the node searches for the  packet's recipient address in its neighbourhood within two-layer depth.  If the recipient is not in the searched area, the  packet is sent to a random neighbour, who repeats the search in its neighbourhood and so on. 
\item {\it Queuing.} When more than one packet is found at the same node, the packets make a queue in the buffer at that node, waiting to be processed. 
We use a fixed maximum buffer size $H=1000$ packets at each node. If the buffer of a selected node is full, i.e., as at the jamming threshold,  the packet can not be delivered and waits for a further possibility to be delivered. One packet per time step is processed.
\item {\it Delivery.} 
When the packet arrives to its destination (recipient node) it is removed from the network.  
\end{itemize}
Simulations reveals that the traffic properties depend on the parameters, i.e., posting rate $R$, maximum queue length $H$, the queuing discipline (LIFO), and the search depth ($d=2$ in the case of $nnn$-search), as well as on the structure of the underlying network. Note that the transport of packets with the $nnn$-search differs from the random diffusion (random-walk dynamics) in that, when the recipient node is found within next-neighbourhood of a processing node, the packet  goes directly to the recipient node. Otherwise, the packet performs a random walk. In this way, the traffic on networks with structure which is more suitable for the $nnn$-search is much more effective. In particular, it was shown in Refs. \cite{tadic2004} that the clustered scale-free network of the structure of Webgraph (i.e., with two hub nodes and a large number of triangles) is most suitable for next-neighbour-search navigation. The statistical features of the traffic on the Webgraph structure (with parameter $\alpha =0.25$) compared to the scale-free tree (see details in \cite{tadic2005}) show, among other measures, shorter travel and waiting times of packets, and up to 40 times larger packet density before the jamming point is reached. These features are important for understanding the traffic on our modular networks shown in Fig.\ \ref{networks}, where, as explained above, the internal structure of each module in Net269 Fig.\ \ref{networks}(top) has the Webgraph structure (with $\alpha =0.9$) and the supporting network in Fig.\ \ref{networks}(bottom) is a scale-free tree.

\begin{figure*}[htb]
\begin{center}
\begin{tabular}{cc}
\resizebox{20.6pc}{!}{\includegraphics{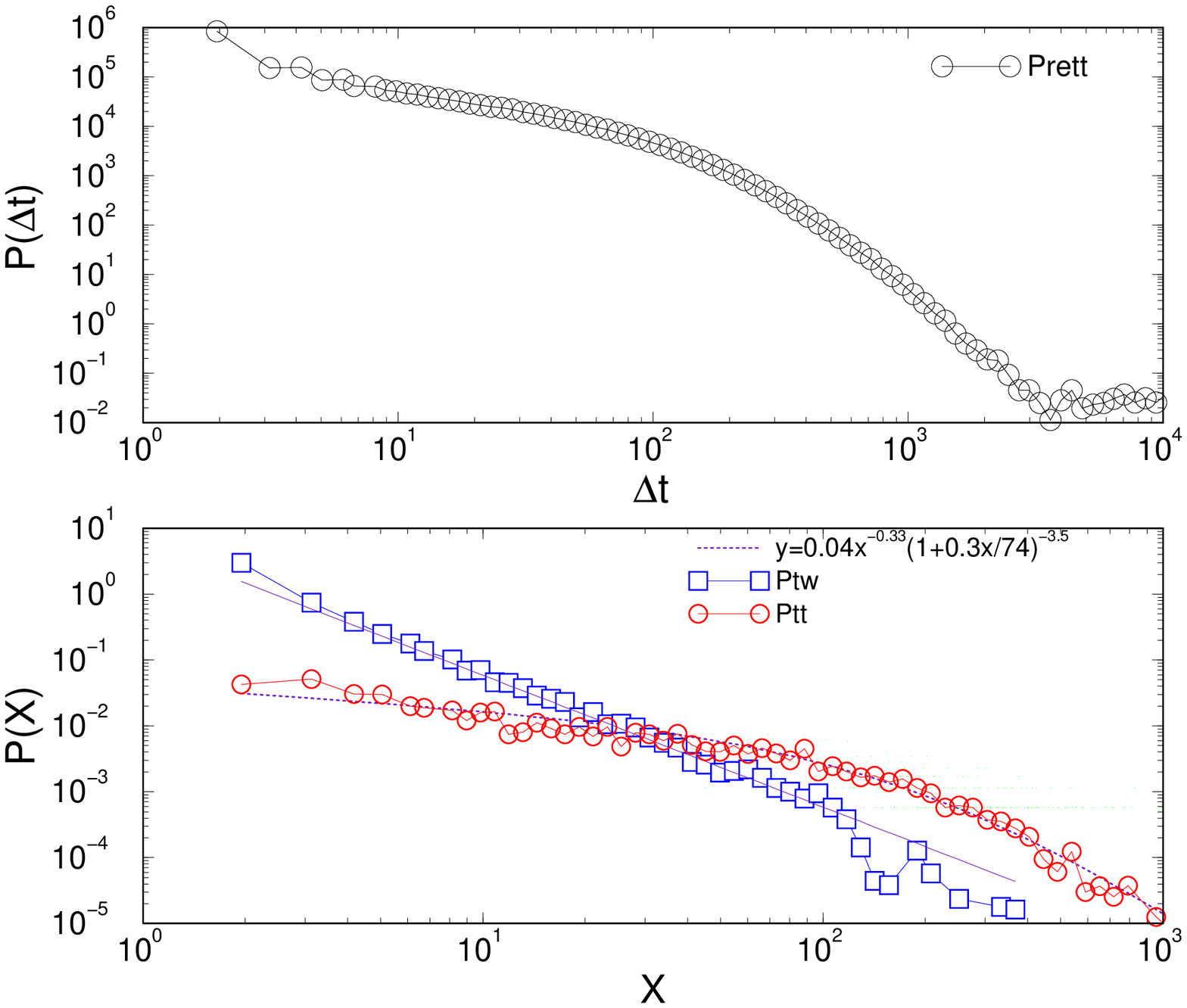}} & 
\resizebox{20.6pc}{!}{\includegraphics{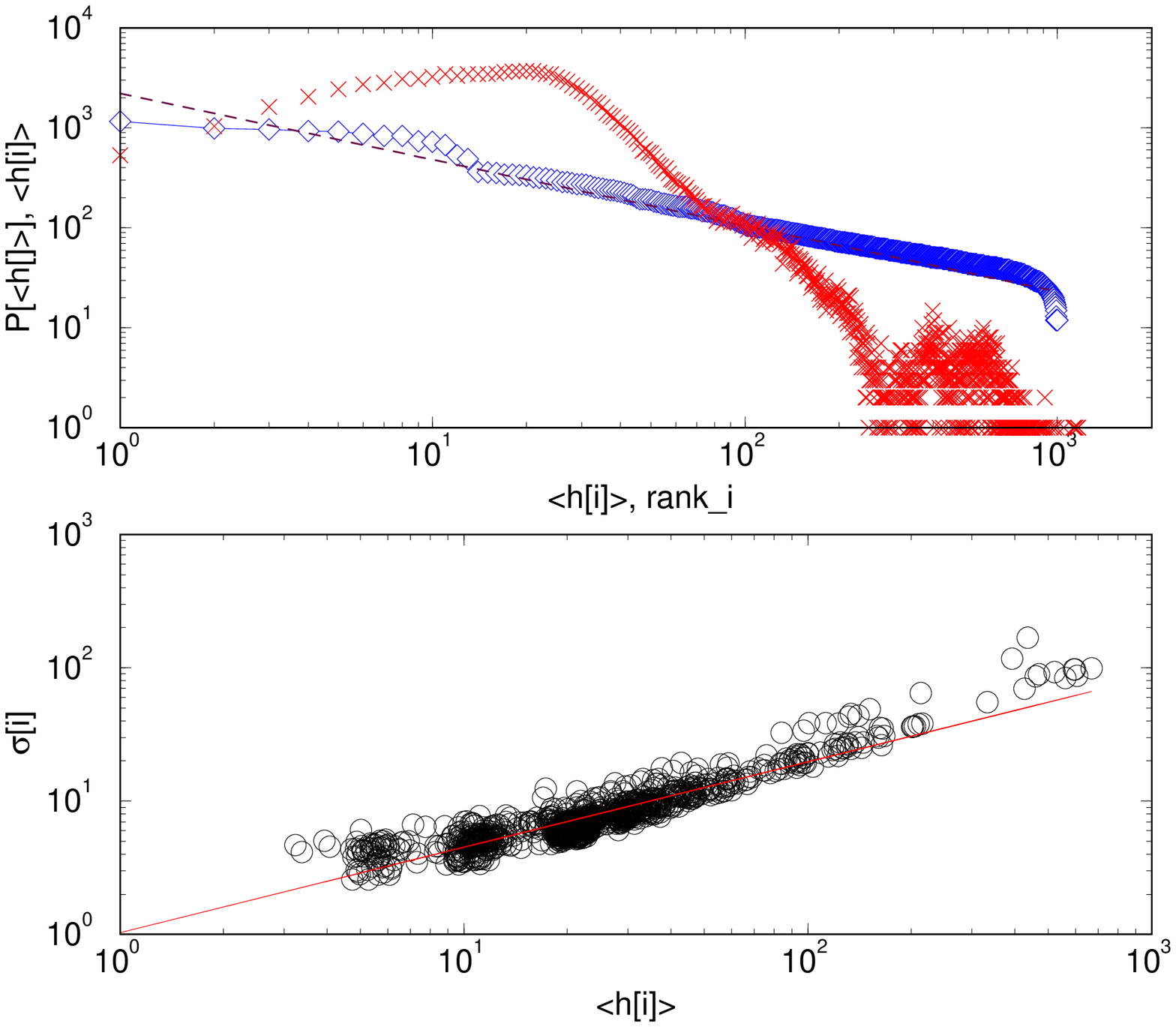}} \\
{\large (a)}& {\large (b)}\\
\end{tabular}
\end{center}
\caption{(a) Distributions of the travel times (Ptt) and waiting times (Ptw) of packets, (bottom panel), and the distribution of  time intervals between successive activity of a node averaged over the network, (top panel). (b) Scatter plot dispersion $\sigma[i]$ vs. average $<h[i]>$ of the time series of all nodes, (bottom), and the ranking distribution of nodes according to the average number of packets $<h[i]>$ ($\diamond$) and the distribution of node occupation within time window $T_{WIN}=10^3$ steps (x).  All data on the Net269 (Fig.\ \ref{networks}top) for posting rate $R=0.8$.}
\label{fig:traffic}
\end{figure*}

In Fig.\ \ref{fig:traffic}(a) bottom panel, we show several results of the traffic on the modular network  Net269.  Measured are  {\it local} and {\it global} statistical properties for fixed posting rate $R=0.8$. The  probability distributions of travel times of packets $P(T_T)$ between creation and delivery and the distribution of waiting times of packets at different nodes along the path, $P(t_w)$ are broad range distributions, which are characteristic for the networks structure and given packet density (below jamming point).  Specifically, the waiting time distribution shows the power-law decay $P(t_w) \sim t_w^{-\tau_w}$, with slope $\tau_w\approx 2.1$, indicating that that at this posting rate $R$ the network operates close to the threshold of jamming (divergence of the average waiting time, which is compatible with 
$\tau_w < 2$, is one of striking features of the traffic jamming). The distribution of travel times appears to be affected with the network modularity. In contrast to the Levy-type distribution on the Webgraph (see \cite{tadic2005}), the  modules of the same structure interconnected as in Fig.\ \ref{networks}(top), lead to a distribution with weak power-law at small times, and a q-exponential tail (see recent work \cite{pluchino2008} and references therein): 
\begin{equation}
P(X) = AX^{-\tau }\times (1 +(1-q)X/X_0)^{1/(1-q)} \ ;
\label{tt-qexp}
\end{equation}
with $\tau \approx 0.33$ and $q=1.29$. A similar expression with two slopes ($\tau \sim 1$ and $q\sim 1.21$) can fit the distribution $P(\Delta t)$ of time intervals between successive events at a node (fit not shown). Note that in the case of dense traffic that we simulate here, the distribution $P(\Delta t)$ refers to 
return of the activity (not the packet) to the same node, and thus it is different from the familiar autocorrelator, the return time of the random walk to the origin  (see \cite{mmbt2008} for the simulations of the autocorrelator). The appearance of two different slopes in the distribution of return of the activity to the node $P(\Delta t)$ in Fig.\ \ref{fig:traffic}(a), suggests uneven role of different nodes in the traffic. This is further studied in terms of {\it number of packets}  $\{h_i(t_k)\}$ processed by each node within a given time window $T_{WIN}=1000$ time steps, shown in Fig.\ \ref{fig:traffic}(b). The time series  $\{h_i(t_k)\}$, where $t_k$ stands for the index of successive time windows, is recorded at each node of the network $i=1,2, \cdots N$. Dispersion of each of these time series, $\sigma [i]$ is plotted in Fig.\ \ref{fig:traffic}(b) bottom panel, against its average value $<h[i]>$. In this scatter plot each point stands for one node of the network. The plot shows the scaling behavior 
\begin{equation}
\sigma [i] = const \times <h[i]>^\mu \ ;
\label{sig-h}
\end{equation}   
with $\mu \in [1/2,1]$, which is characteristic for many real dynamical systems (see recent review \cite{esler2008}). 
\begin{figure*}[htb]
\begin{center}
\begin{tabular}{cc} 
\resizebox{18pc}{!}{\includegraphics{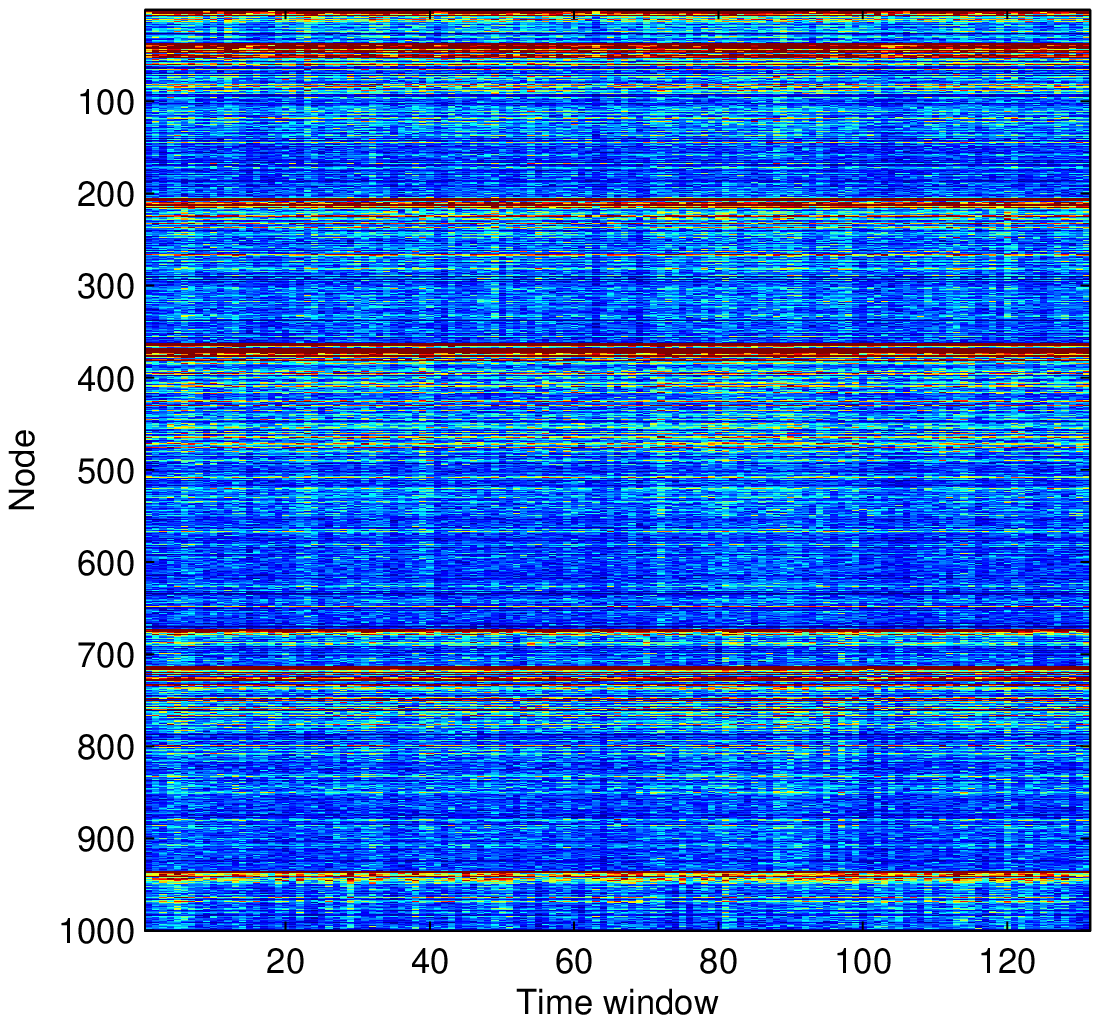}} &
\resizebox{18pc}{!}{\includegraphics{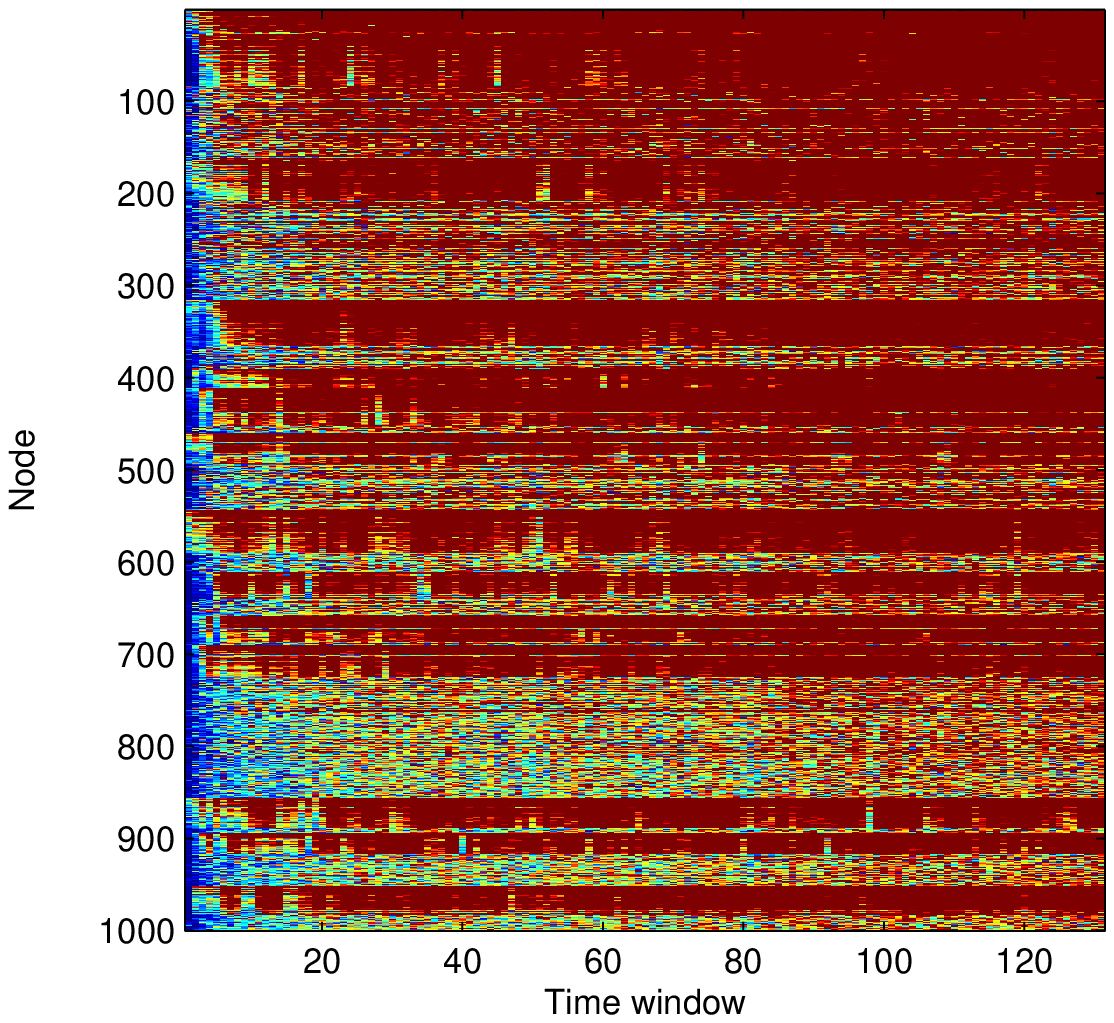}}   \\
{\large (a)}& {\large (b)}\\
\end{tabular}
\end{center}
\caption{Three-dimensional plot of the time series of the number of packets processed by a node within a time window $T_{WIN}=10^3$ steps, $\{h_i(t_k)\}$, for all nodes in the network $i=1,2, \cdots 1000$ and for 132 time windows for Net269 (a) and for tree-with-attached modules (b). Color map: dark-blue corresponds to low, while dark-red to highest recorded value. }
\label{fig:3dplot}
\end{figure*} 

The origin of such scaling in traffic models on complex networks has been attributed \cite{ktr-njp07} to node groupings according to their role in the traffic, which, in turn, is related to their topological (or dynamical) centrality. At the tip of the plot in Fig.\ \ref{fig:traffic}(b) bottom, one can distinguish a group of 10 most active nodes. The same group is also separated in the ranking plot (flat part at the beginning of the curve in the upper panel) from the rest of nodes. The ranking plot exhibits a power-law decay (Zipf's law) according to 
\begin{equation}  
<h[i]> \sim rank_i^{-\gamma} \ ;
\label{eq:ranking}
\end{equation}
 with $\gamma \approx 1$, where ranking $rank_i$ of nodes $i=1,2, \cdots N$ is made according to the average number of packets  $<h[i]>$ processed by the node $i$.  
Furthermore uneven role of nodes is demonstrated in the probability distribution of the number of packets processed within the time window $<h[i]>$ (or the ``occupation probability'' of a node $\rho [i] \equiv <h[i]>/T_{WIN}$). The distribution is also shown in Fig.\ \ref{fig:traffic}(b) top panel.  The central peak appears resembling the ``ergodic'' system behavior \cite{barkai2007}, however, at the side of large occupancy additional peaks are present representing the activity of modular hubs and other highly active nodes within the modules. Note that for this traffic density (posting rate) there the nodes which have maximum occupancy $\rho =1$ may occur, which is another signature of the pre-jamming behavior in the network.

The activity of different nodes and their role in the traffic process on the modular network Net269 are shown comparatively in Fig.\ \ref{fig:3dplot}(a). The most active nodes (red) are modular hubs. Gradually lesser activity is visible  at nodes within each module, according to their reduced centrality (scale-freeness of the internal module structure). For the comparison, the 
3-dimensional plot of the time series of all nodes in the scale-free tree with attached modules is given in Fig.\ \ref{fig:3dplot}(b). Jamming on the tree graph (indicated in red color) is visible at the place of modules. This indicates that  packets  often get confined within a module due to ineffective search mechanism on the underlying tree graph. In the following section we analyse the correlations between these time series and the spectra of the correlation matrix.

\section{Correlations and Spectra} \label{sec:correl}

\subsection{Correlation of traffic time series}

\begin{figure*}[htb]
\begin{center}
\begin{tabular}{cc} 
{\large (a)}&{\large (b)}\\
\resizebox{16.8pc}{!}{\includegraphics{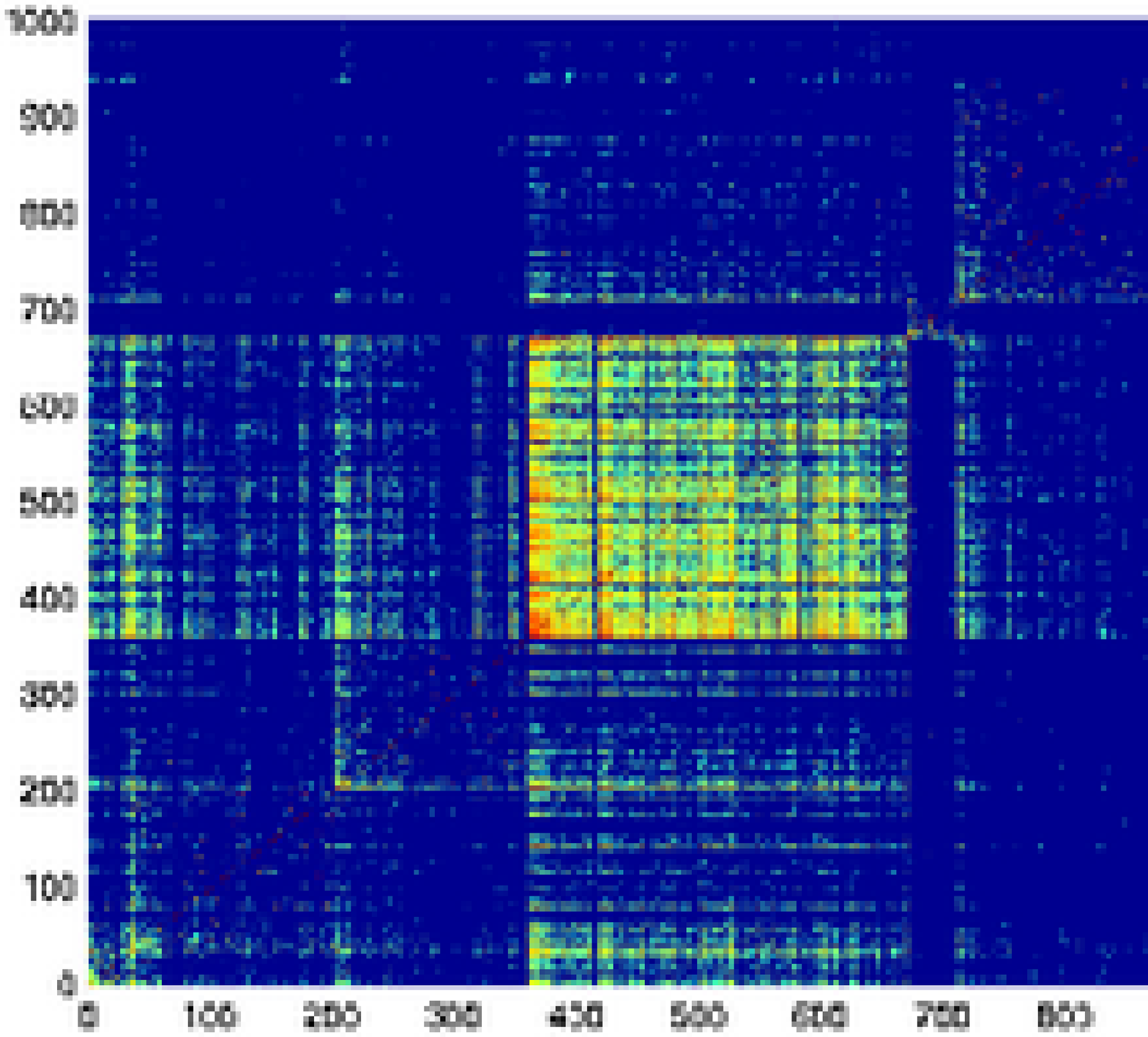}}&
\resizebox{16.8pc}{!}{\includegraphics{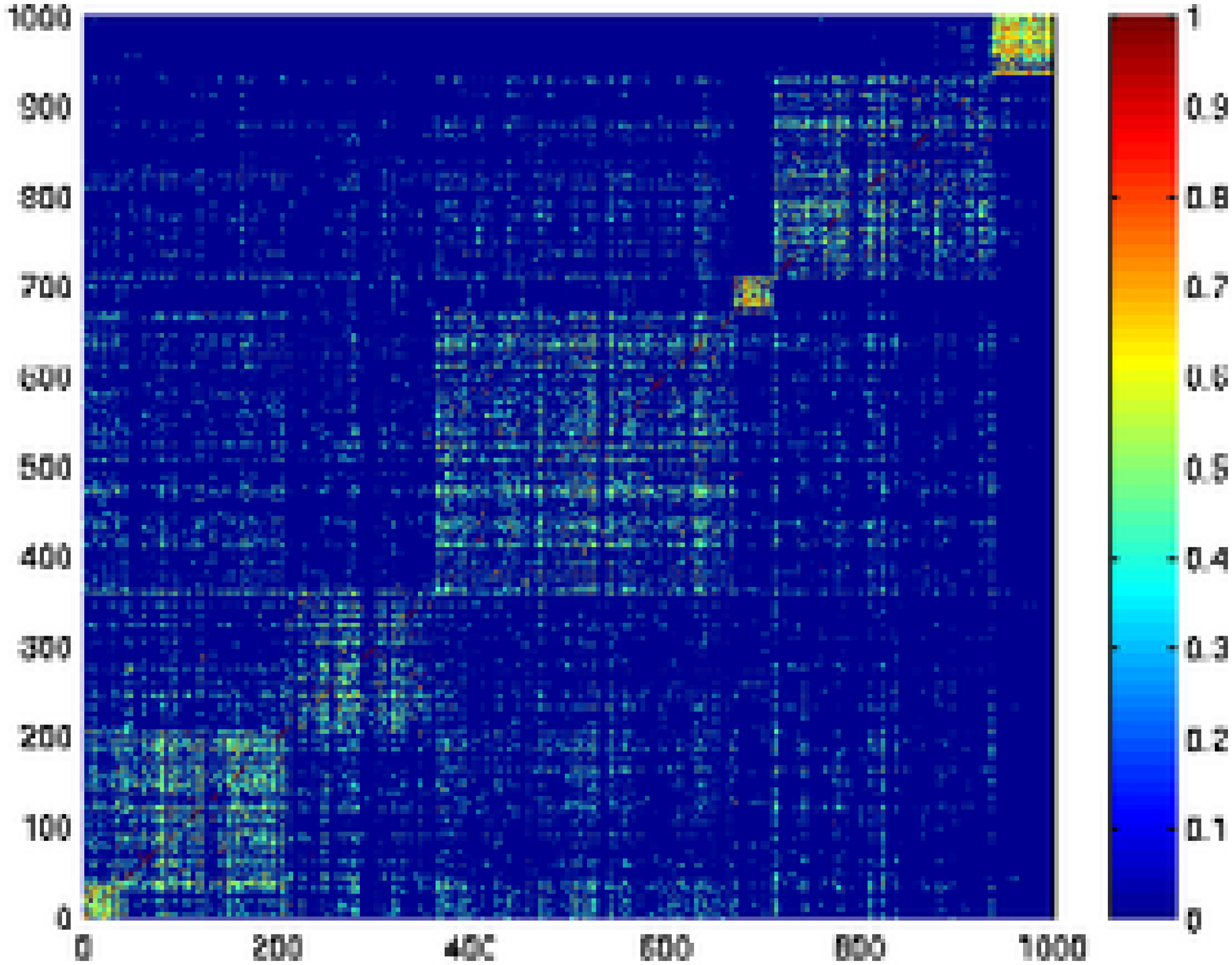}}\\
\resizebox{16.8pc}{!}{\includegraphics{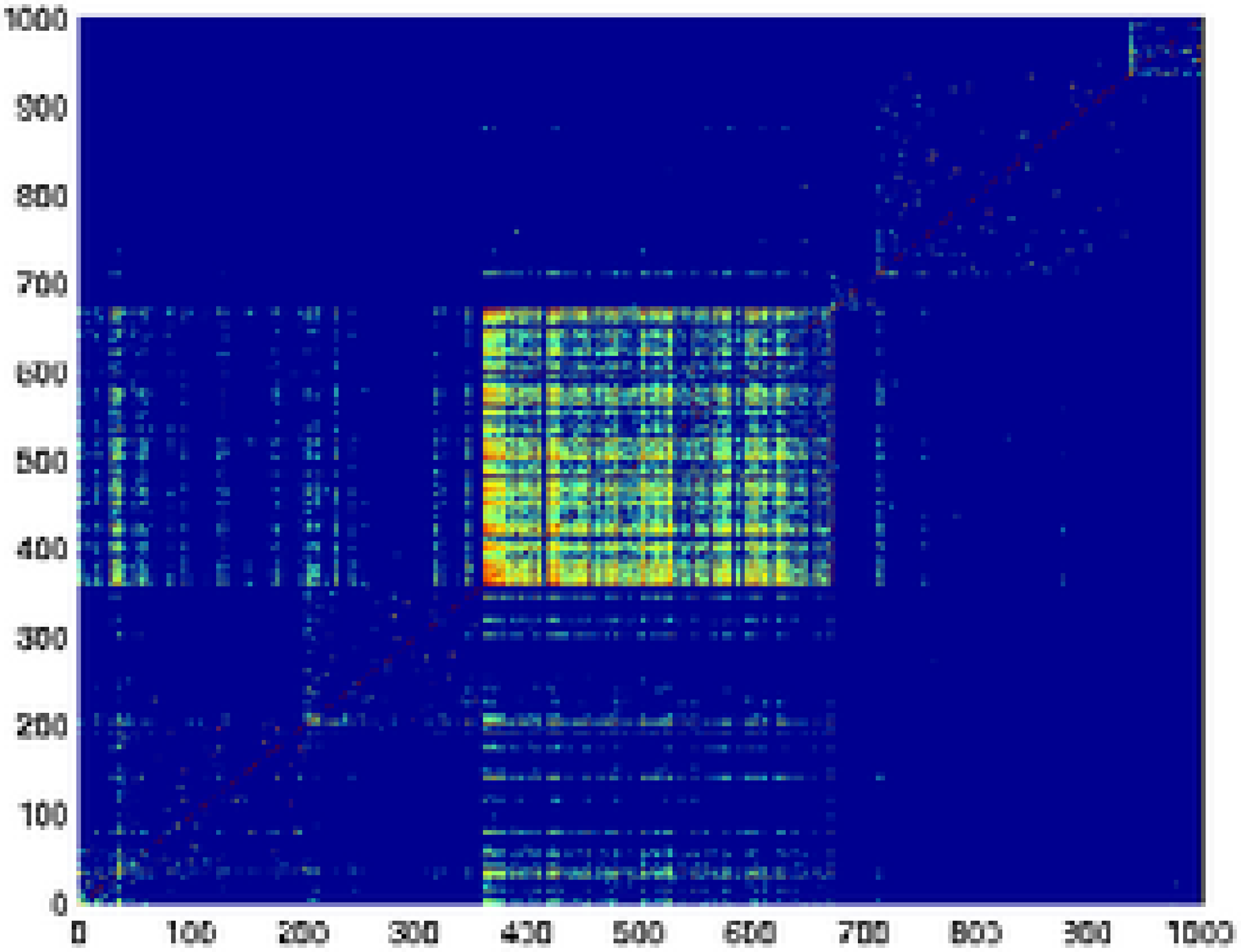}}&
\resizebox{16.8pc}{!}{\includegraphics{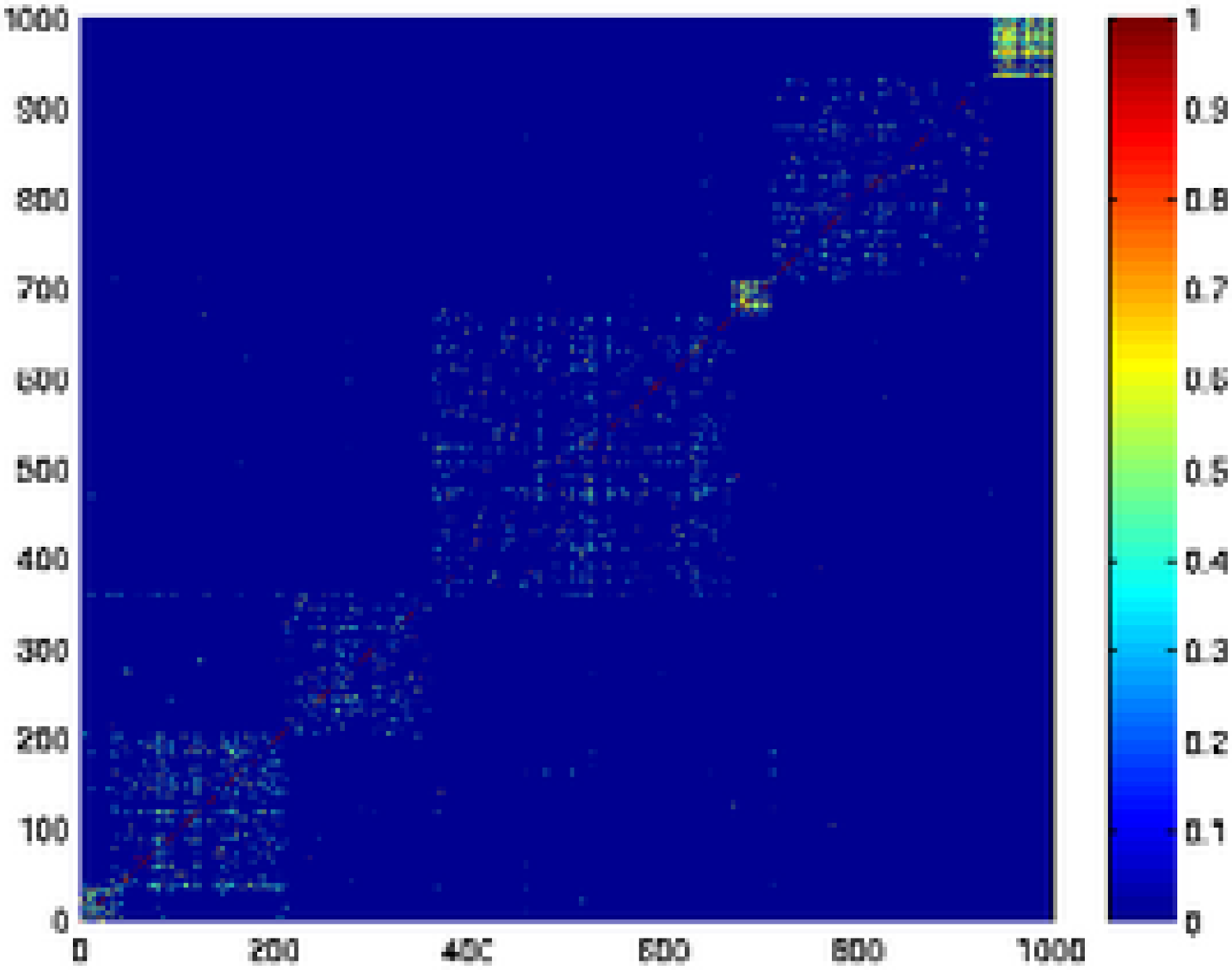}}\\
{\large (c)}&{\large (d)}\\
\end{tabular}
\end{center}
\caption{Correlation matrix of the traffic signals on  (a) Net269  and (b) on same network with randomized links inside the modules. (c) and (d): Corresponding filtered correlation matrices. In all cases shown are links above the threshold value  $C_{o}=0.4$.}
\label{corr_net}
\end{figure*}

\begin{figure}[htb]
\begin{center}
\begin{tabular}{c} 
\resizebox{21pc}{!}{\includegraphics{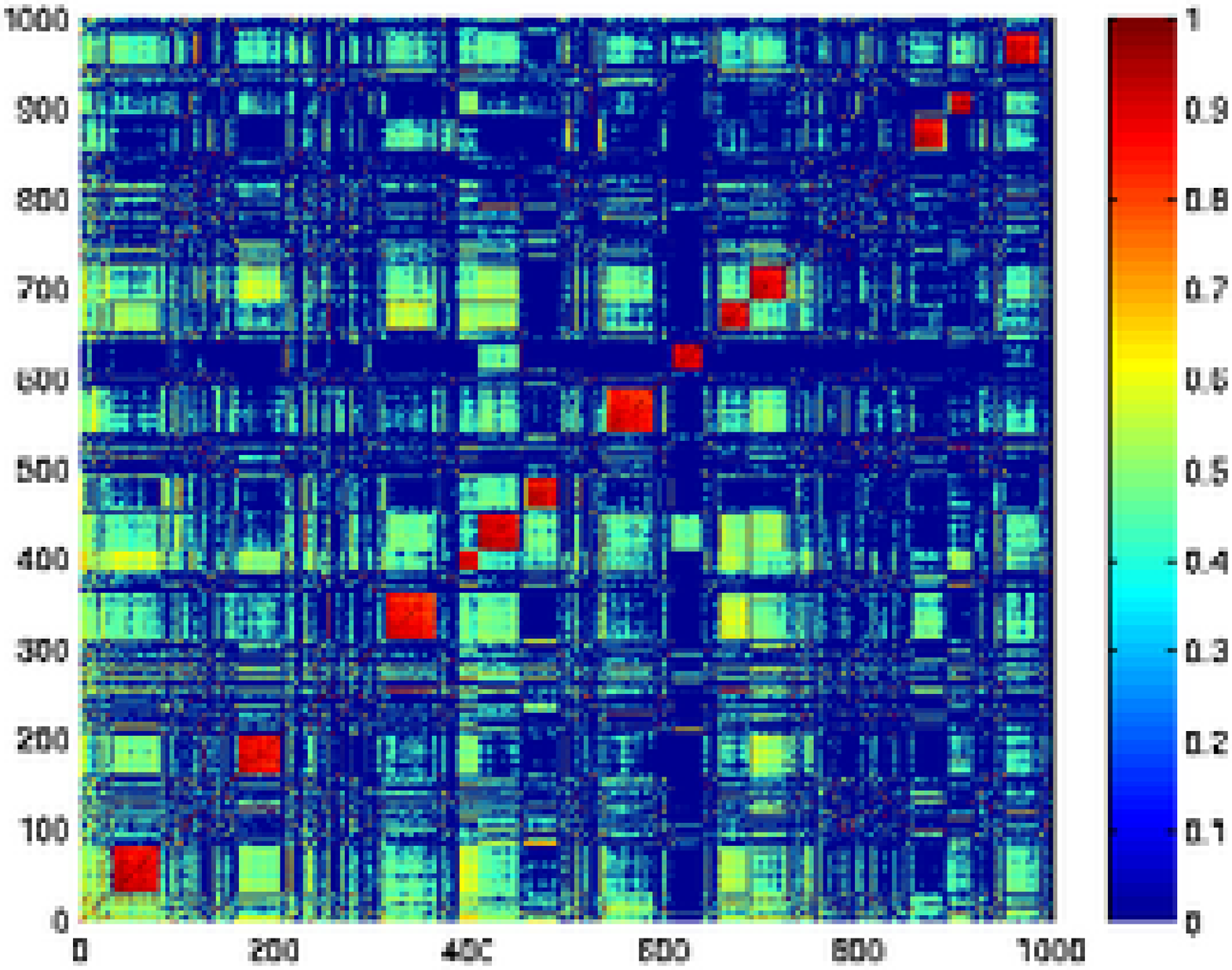}}\\
\resizebox{21pc}{!}{\includegraphics{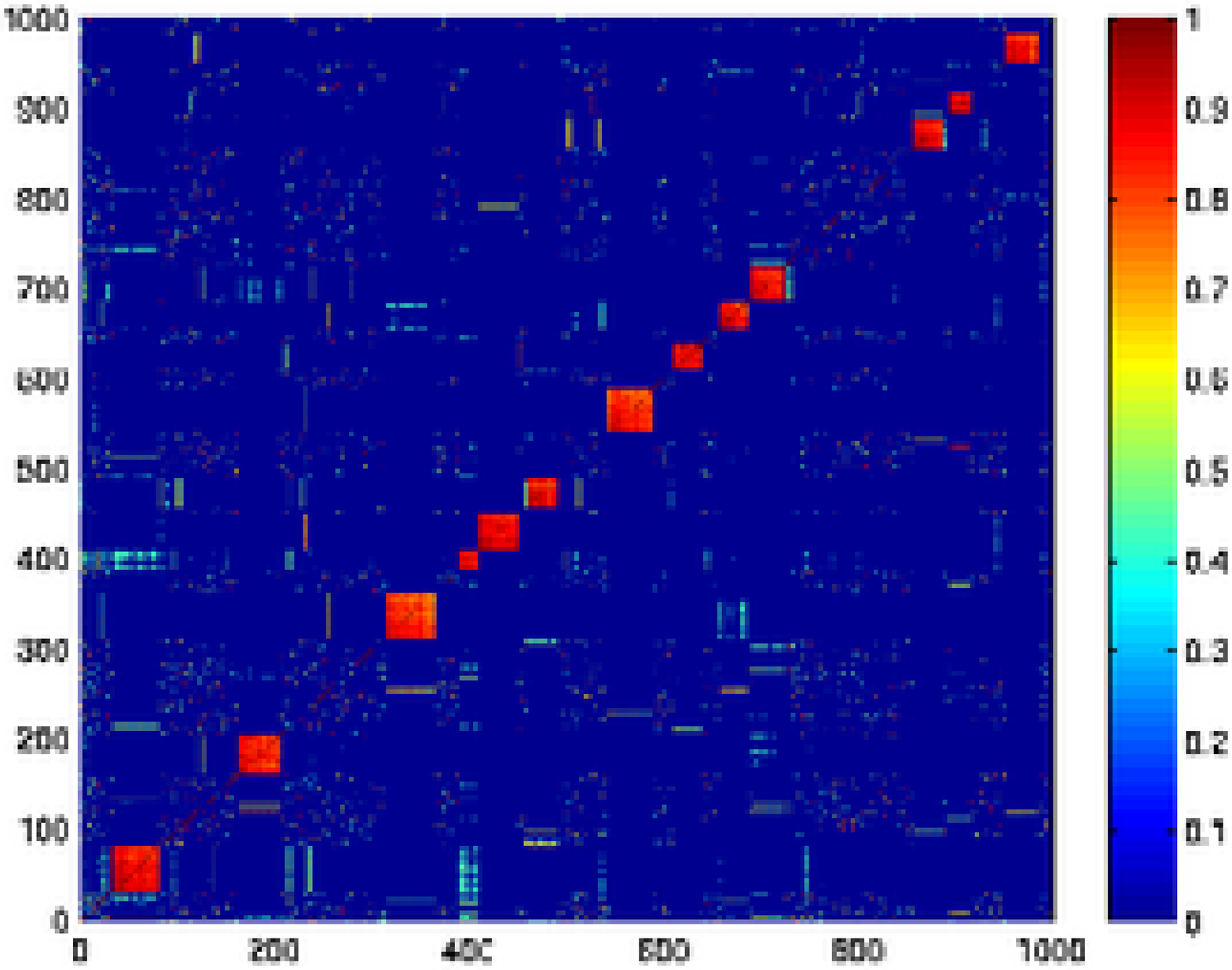}}\\
\end{tabular}
\end{center}
\caption { Correlation matrix (top) and filtered correlation matrix (bottom) obtained from the traffic time series on tree with attached modules. Shown are the links above the threshold value $C_0=0.4$ . }
\label{corr_net_trepp}
\end{figure}

The elements of the correlation matrix $C_{ij}$ are  obtained by calculating the Pearson's correlation coefficient between time signals $h_{i}(t_{k})$ for each pair $i,j$ of nodes in the network, given by:
\begin{equation}
 C_{ij}=\frac{\sum_{t_{k}}[h_{i}(t_{k})-<h_{i}>][h_{j}(t_{k})-<h_{j}>]}{\sigma_{i}\sigma_{j}} \ . \label{pearson}
\end{equation}
Here $h_{i}(t_{k})$ is activity of node $i$ in time-window $t_{k}$, $<h_{i}>$ is average activity during the whole time period,
\begin{equation}
\langle h_{i} \rangle=\frac{1}{n_{t}}\sum_{t_{k}}h_{i}(t_{k})  \ , 
\end{equation}
where $n_{t}$ is the number of {\it time windows} considered, and $\sigma_{i}$ is standard deviation of the time signal at node $i$:
\begin{equation}
\sigma^{2}_{i}=\frac{\sum_{t_{k}}(<h_{i}>-h_{i}(t_{k}))^{2}}{n_t} \ .
\end{equation}
The Pearson correlation coefficient takes values from $-1$ (strong anti-correlations) to $+1$ (strong correlations between nodes).
Using Eq.\ (\ref{pearson}) we obtain the correlations between nodes in modular networks from the time signals obtained in simulations of traffic, presented in Fig.\ \ref{fig:3dplot}. For this kind of time signals, the distribution of the correlation coefficients $P(C_{ij})$ strongly depends on the overall traffic density (or posting rate $R$), and for the posting rate approaching the jamming threshold used in this simulations, the peak of the distribution is moved towards right edge. Specifically, for the signals shown in Fig.\ \ref{fig:3dplot}, the  
values of correlations are centered around $c=0.2$ (Net269) as to $c=0.35$ (tree with attached modules), with negligible density of the negative correlations. Nodes connected in the original network have high positive correlation coefficient which is a result of their similar activities during the measurement time.
The 3-dimensional plot of the correlation matrix for the traffic signals on the modular network Net269 is shown in Fig.\ \ref{corr_net}(a). Similarly, the correlation matrix of the signals recorded on the tree network with attached modules is shown in Fig.\ \ref{corr_net_trepp}(top).  
It is clear that in both networks, the most active nodes in each module (module hubs) have a high correlation  coefficient with each other and  with the rest of the nodes (cf. Figs. \ref{corr_net} and \ref{corr_net_trepp}), although topological connections between them might not be present.

In order to extract the information about network structure from the correlation matrix, we first observe that only correlations above certain  threshold $C_{o}$, i.e., $C_{ij}>C_{o}$ might be relevant. In this way one attempts to separate the values of  potentially relevant correlations from those which arise accidentally (and are normally distributed around central peak). 
Although many weak correlations are filtered out in this way, the remaining matrix still contains many spurious links, compared with the original adjacency matrix of the Net269. This is demonstrated graphically in Fig.\ \ref{corr_net}(a),  where the  correlation matrix of the traffic signals  is shown with threshold $C_{o}=0.4$. Note that with this threshold the remaining correlation matrix contains a single connected component. Higher threshold values may result in fragmentation. 

As the Fig.\ \ref{corr_net}(a) shows, the correlation matrix contains already information about modules (diagonal blocks) and their size. The picture is much less clear in the case of small modules on the scale-free tree (cf. Fig.\ \ref{corr_net_trepp} (top)). Generally, the hub of the scale-free tree has strong correlations with other nodes, because of the large signal (large number of packets processed) on the hub: first row and first column in the matrix. Similarly, the correlations are enhanced between the hubs of the large modules in Fig.\ \ref{corr_net}(a), since the hubs inside the modules carry the largest traffic. For small modules on the tree, the walker gets trapped inside the module for longer time, since the module is linked  to the rest of the network by a single node. Hence, enhanced correlations between the modules remain above the threshold and are seen as the off-diagonal blocks in Figs.\ \ref{corr_net} and \ref{corr_net_trepp}. 
Generally,  more sophisticated methods are necessary in order to reduce the number of such spurious correlations, which are not related with the occurrence of a direct link between the nodes in the adjacency matrix \cite{muller,eshel2004}.
Here we apply one of the filtering methods which utilizes the affinity transformation \cite{eshel2004,eshel2008}.
The idea is to multiply each element $C_{ij}$ of the correlation matrix with a factor $M_{ij}$ which is constructed from the elements the rows $i$ and $j$ in the correlation matrix in the following way: Excluding the diagonal elements $C_{ii}$ and $C_{jj}$ the remaining matrix elements of the correlation  matrix are first reordered to form the $n\equiv (N-1)$-dimensional vectors $\{C_{ij}, C_{i1},...,C_{in}\}$ and $\{C_{ji},C_{j1},...,C_{jn}\}$. Then $M_{ij}$ is computed as the Pearson's coefficient of the components of these vectors. Then the  matrix element $C^{M}_{ij}$ of the filtered correlation matrix is given by the product
\begin{equation}
C^{M}_{ij}=M_{ij}C_{ij}\ . \label{meta}
\end{equation}
In this way,  the correlation between the  nodes $i$ and $j$ is enhanced if the corresponding  meta-correlation element $M_{ij}$ is large (i.e., the nodes $i$ and $j$ see the rest of the network in a similar way), and reduced otherwise.
 If  two nodes are linked on the network, it is expected that their correlations with other nodes are similar, resulting in value of their meta-correlation close to one, otherwise coefficient $M$ is closer to zero. Hence, the multiplication of the elements of the correlation matrix with meta-correlations should increase the difference between true and random correlations.
For the filtering procedure we use the whole correlation matrix without any threshold. Note also that, in contrast to most general case where a shift of the interval [-1,1] to [0,1] is performed before the filtering, our correlation matrix of traffic signals is naturally shifted to the positive side. The filtered correlation matrix is also shown in Fig.\ \ref{corr_net}(c), only the links  stronger than the threshold are shown. In comparison with the unfiltered correlation matrix, the number of matrix elements is considerably reduced relative to the same threshold value. The effects are much more prominent in the case of the tree with small modules, shown in Fig.\ \ref{corr_net_trepp} (bottom).  

We further notice that the internal structure of the (large) modules play a role in the dynamics and the correlation patterns. The occurrence of hubs, better connected nodes which carry most of the traffic inside the modules, leads to strong correlations of the traffic signals with other  nodes within the module and between different modules. Some of these correlations can not be filtered out, as seen in Fig.\ \ref{corr_net}(c). 
In order to support this conclusion, we made random rewiring of the links inside each module by keeping the total number of links and links between the modules unchanged. Then we run the traffic on the randomized network and construct the correlation matrix of the traffic signals. The results are shown in Fig.\ \ref{corr_net} (b) and (d): the correlation matrix of the randomized network and the corresponding filtered correlation matrix (lower panel), suggesting that the filtering procedure is much more effective in the case of randomized structure of the modules.

By applying the threshold in the filtered correlation matrix one can visualize the graph structure (shown in Fig.\ \ref{ranking_spectra_ad_corr}, top) for both original and randomized version of modules. Compared to the original adjacency matrix, these structures contain many spurious links, although the number of such links is considerably smaller in the randomized version. It is interesting to note that the association of the nodes to given modules is almost entirely preserved as in the original network. In Fig. \ref{ranking_spectra_ad_corr} the links represent the filtered correlation matrices, but each node carry the color which indicates its membership to a module in the original network Net269. Again, the randomized connection inside the modules lead to the correlation network with clear modular structure with only few nodes wrongly attributed to their original modules.

\subsection{Spectral analysis \label{sec:correl-spectra}}
Further details of the interdependences between the diffusion on networks and their structure are obtained through the spectral analysis of the Laplacian operator. The detailed spectral analysis of the normalized Laplacian related to the modular networks  has been reported in Ref.\ \cite{mmbt2008}. The spectral density and the structure of the corresponding eigenvectors show specific features which are related to the modularity and other properties (clustering, average connectivity, etc.) of the networks.  

Here we perform spectral analysis  of the normalized Laplacian related to the correlation  matrices $\mathbf{C}$ and $\mathbf{C^{M}}$ for both types of the modular network structures discussed above. 
In order two exclude selfedges, all elements on diagonal are set up to zero value. The normalized Laplacian related to random-walk type dynamics is given by \cite{mmbt2008,samukhin2007}
\begin{equation}
L_{ij}=\delta_{ij}-\frac{A_{ij}}{\sqrt{q_{i}q_{j}}} \ , \label{lap3}
\end{equation}
where $A_{ij}$ are elements of adjacency matrix of the graph, and $q_{i}$ and $q_j$ are  the degree of node $i$ and $j$. It has limited spectrum in the range $[0,2]$ and an orthogonal set of eigenvectors, which makes it suitable for numerical study and comparisons of different structures. For the networks extracted from the correlation matrix $\mathbf{C}$ and filtered correlation matrix  $\mathbf{C^{M}}$ the Laplacian is obtained by constructing the {\it binary graph} $\mathbf{A}$ in Eq. (\ref{lap3}) with the elements of the matrix $\mathbf{C}$ or $\mathbf{C^{M}}$, where $C_{ij}>C_0$ are set to unity, and zero otherwise. As we show in our previous work \cite{mmbt2008}, the spectral properties of the Laplacian depend on the network topology, and can be used for the identification of its mesoscopic structure.
\begin{figure*}[htb]
\begin{center}
\begin{tabular}{cc} 
{\large (a)}& {\large (b)}\\
\resizebox{16pc}{!}{\includegraphics{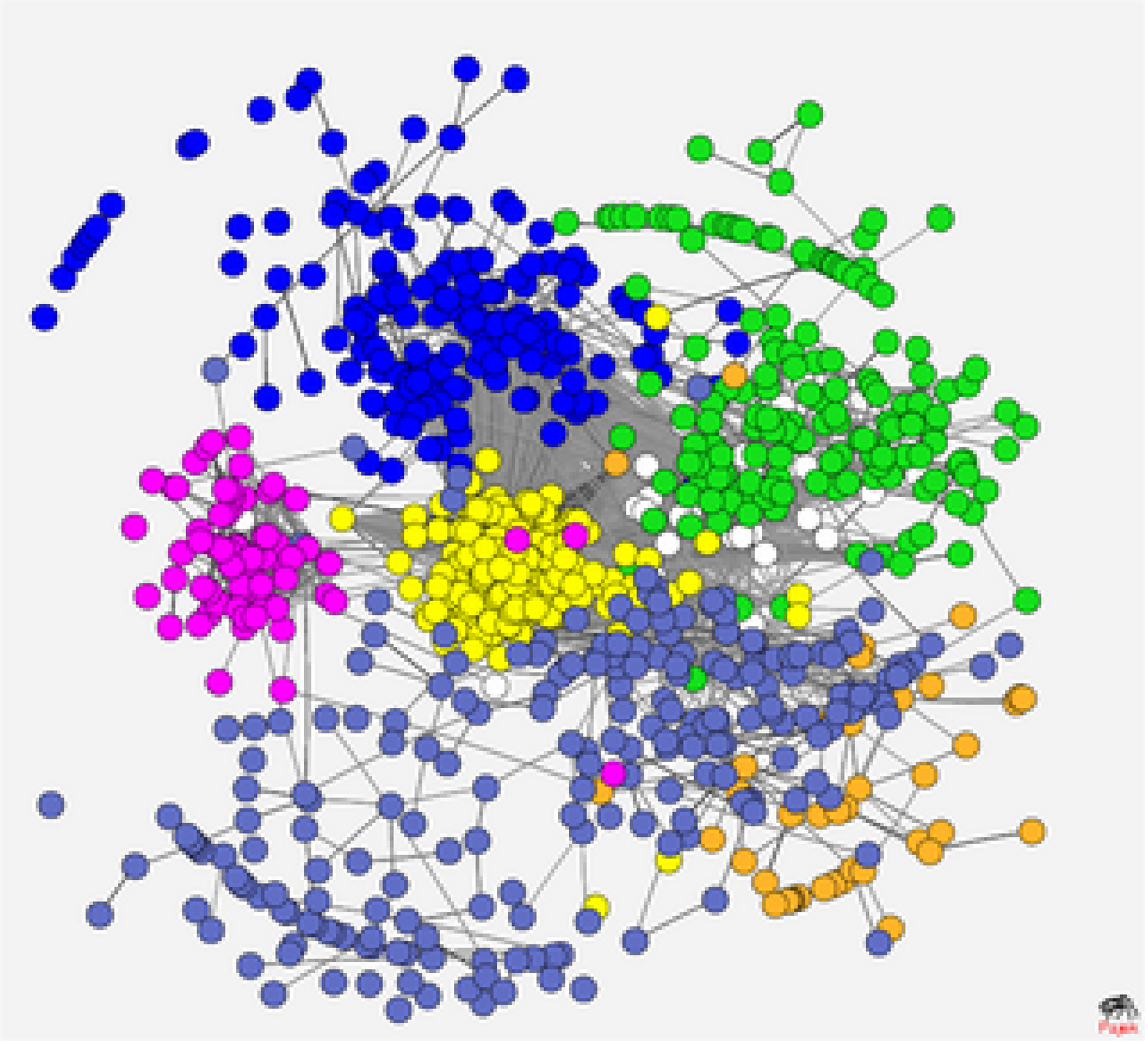}}&
\resizebox{16pc}{!}{\includegraphics{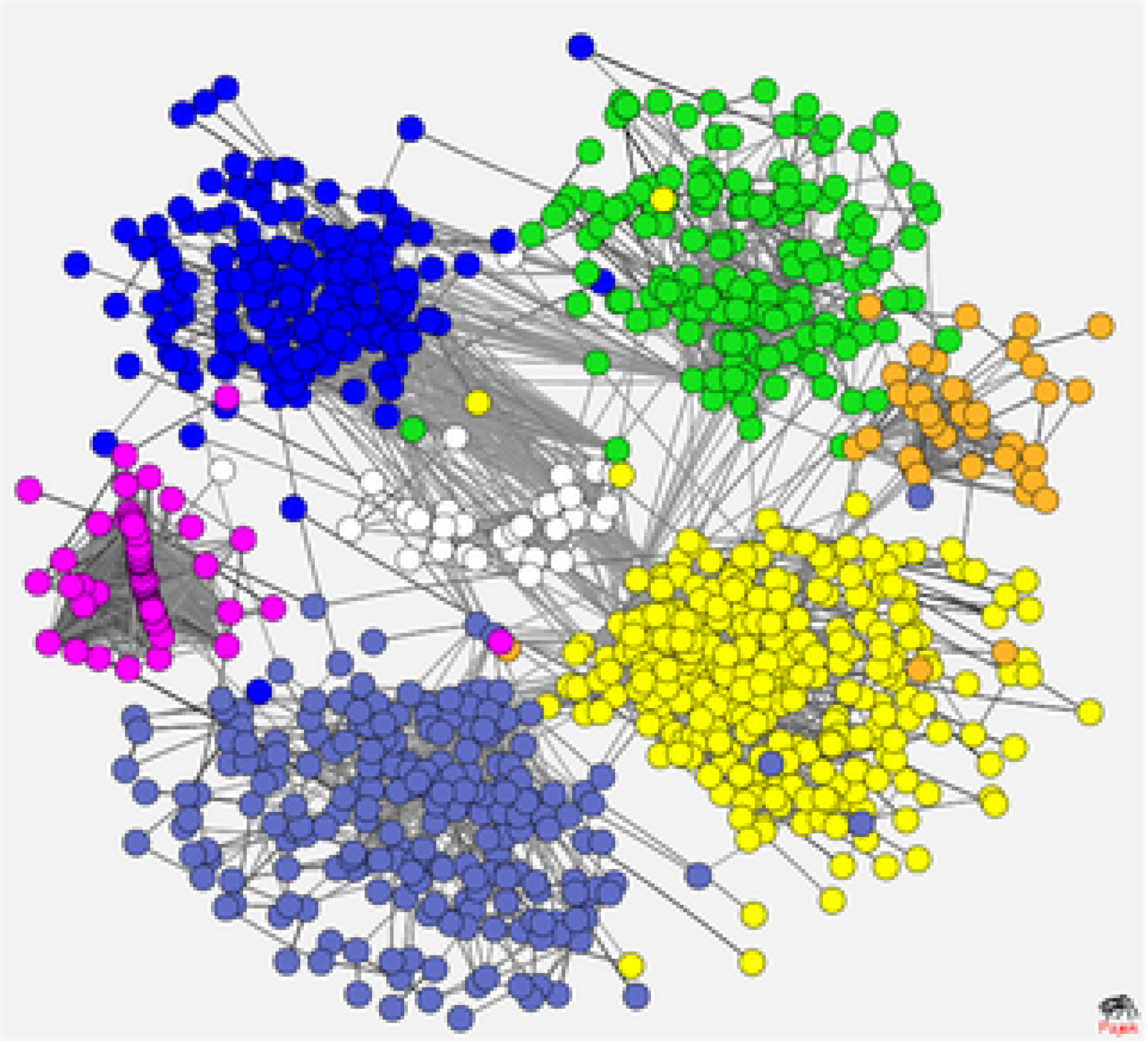}}\\
\resizebox{20pc}{!}{\includegraphics{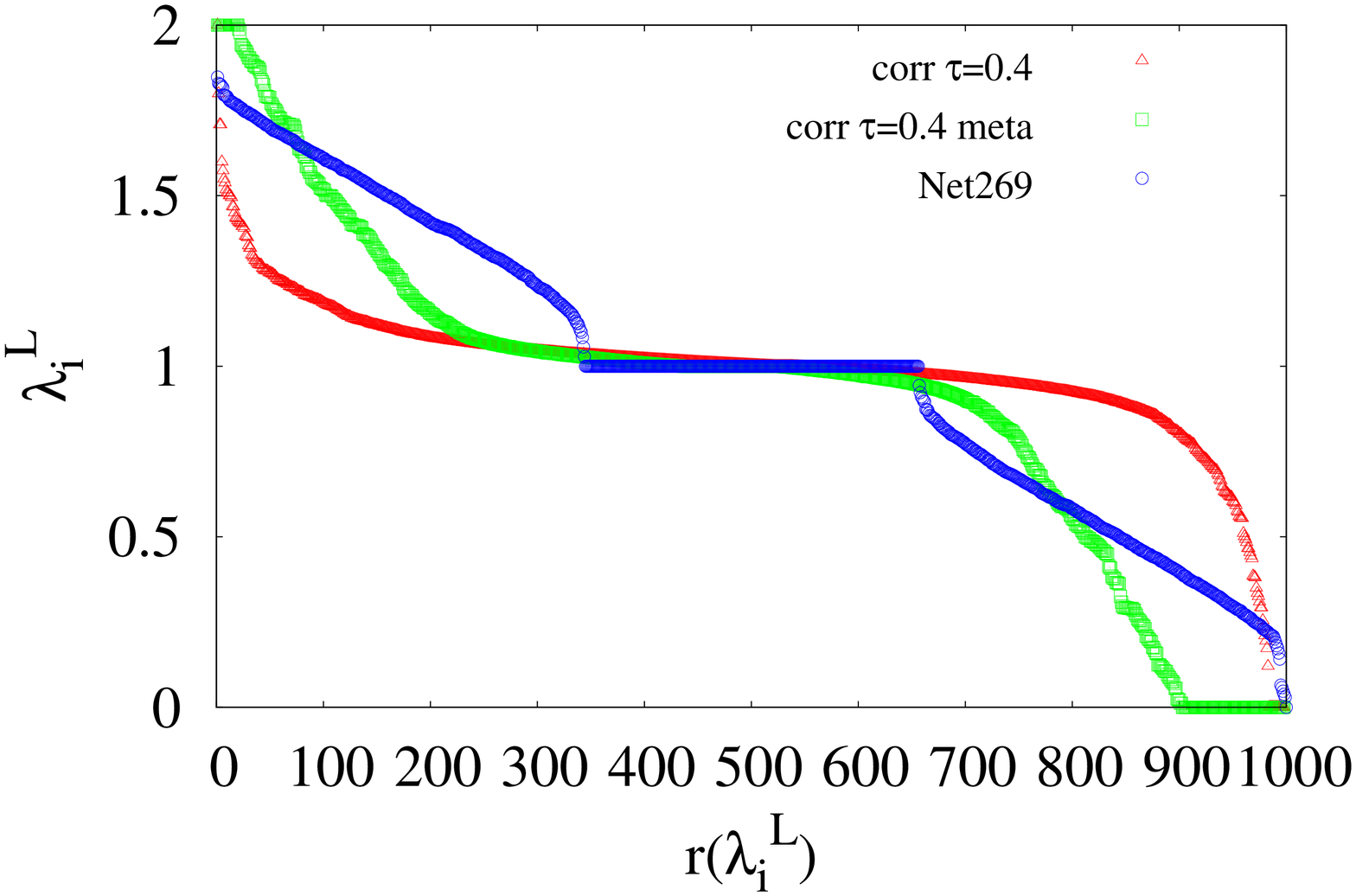}}&
\resizebox{20pc}{!}{\includegraphics{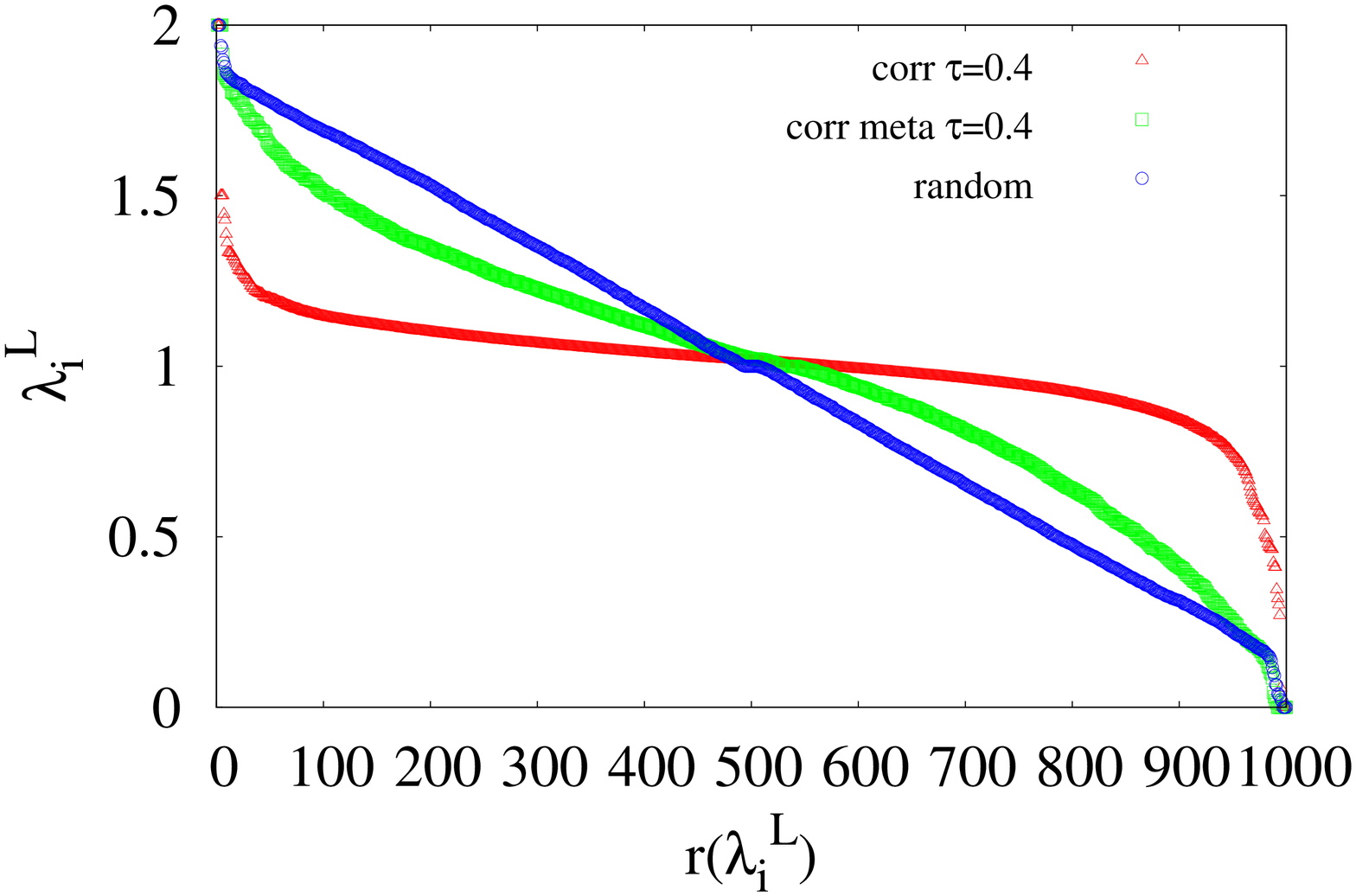}}\\
{\large (c)}& {\large (d)}\\
\end{tabular}
\end{center}
\caption{ Weighted networks obtained from filtered correlation matrices for traffic on  Net269 (a) and on  network with randomized module structure (b).  Shown are links above the threshold $C_0=0.4$. Colors indicate the original membership of nodes to different modules of the Net269. Ranking of eigenvalues of the Laplacian matrix for networks obtained from correlation matrix and filtered  correlation matrix compared to one obtained for original network Net269 (c) and randomized network (d). }
\label{ranking_spectra_ad_corr}
\end{figure*} 

The spectrum of the Laplacian related to the original Net269 has six lowest nonzero eigenvalues, Fig. \ref{ranking_spectra_ad_corr}(c), which are separated from the rest of the spectrum,  and the largest eigenvalue $\lambda_{max} <2$. 
The number of smallest eigenvalues ($\lambda\gtrsim 0$) are correlated with number of well separated subgraphs in the network, while the number of eigenvalues $\lambda=0$ corresponds to number of disconnected components \cite{mmbt2008}. 
The spectrum of the networks obtained from $\mathbf{C}$, also shown in  Fig.\ \ref{ranking_spectra_ad_corr}(c), has $15$ zero eigenvalues or disconnected components for the applied threshold $C_0$.  The maximal value $\lambda^{L_{C}}=2$ also indicate  that a subgraph with tree-like structure (chain of nodes) \cite{mmbt2008,jost2008} occurs, in contrast to the original network Net269. 
The number of single nodes or disjointed subgraphs containing a few nodes in the network increase after filtering, see Fig. \ref{ranking_spectra_ad_corr}. Furthermore,  meta-correlation network has continuous spectrum up to zero, although the network exhibits modular structure. The absence of the gap between the lowest eigenvalues and the rest of the spectrum, which is characteristic for modular networks,  is due to many spurious links between hubs which increase connectivity between modules, as also seen in Fig.\ \ref{corr_net}(a). The high density of links, and thus higher average connectivity in correlation networks,  affects the spectrum in the middle part around  $\lambda^{L}=1$, shown in Fig.\ \ref{ranking_spectra_ad_corr}(c). 
In the case of the network with randomized links in the modules, shown in Fig.\ \ref{ranking_spectra_ad_corr}(b), the situation is more clear: the spectra with the filtered correlation matrix and the real adjacency matrix  coincide at both ends, which is compatible with the precise structure of the modules in the top panel of the same Figure. In the intermediate part, however, the deviation between the filtered and real adjacency matrix are large, suggesting that this part of the spectrum can not be effectively used for the identification of the true network structure.
In the case of the scale-free tree with attached small modules the analysis of the spectra of the filtered correlation matrix is entirely ineffective (not shown), suggesting that other methods are necessary for the identification of their modular structure from the dynamical time series.

\begin{figure}[htb]
\begin{center}
\begin{tabular}{cc} 
\resizebox{20pc}{!}{\includegraphics{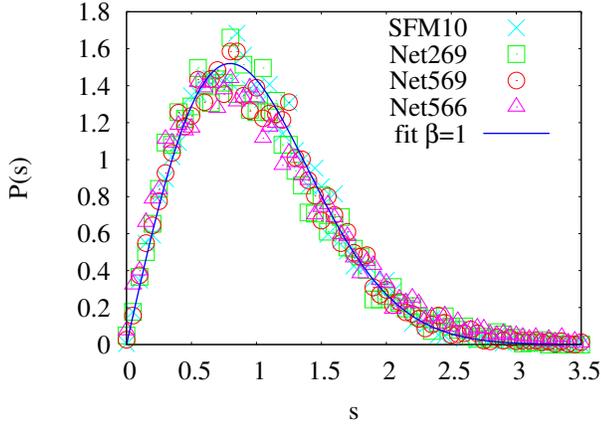}}\\
\end{tabular}
\end{center}
\caption{Distribution of eigenvalues spacing $s_{i}=\lambda_{i+1}-\lambda_{i}$ for non-modular network $M=10,\alpha=1$ and for modular networks with different average connectivity and different value of clustering coefficient: Net269 ($M=2,\alpha=0.9$, $P_{o}=0.006$), Net569 ($M=5,\alpha=0.9$, $P_{o}=0.006$) and Net566 ($M=5,\alpha=0.6$, $P_{o}=0.006$)}
\label{brody}
\end{figure} 

\textit{Spacing distribution} is another feature the spectra which is often studied in the context of the random matrix theory.    Recently it was shown that the  distribution of nearest-neighbor spacing  of the eigenvalues for various model networks (i.e., small world, random networks and scale free networks), follow the universal Gaussian orthogonal ensemble (GOE) of random matrix theory \cite{sarika}.
In this part of the paper we study the spacing of the eigenvalues in various modular networks. We show that, although the spectra of the modular networks exhibit characteristic features related to their mesoscopic structure, connectivity and clustering (see \cite{mmbt2008}), the spacing distribution of their spectra can not be distinguished from the ones in the random matrices.
In Fig.  \ref{brody} the distributions for the spacing of the eigenvalues of the {\it adjacency matrices} of different modular networks with varied structural parameters are shown.
We calculate nearest neighbor spacing distribution for modular networks with different average connectivity, $M$, and different clustering coefficient, grown according to the algorithms in \cite{mmbt2008}, and compare it with non-clustered, non-modular scale free networks. 

We use the standard procedure to determine the nearest-neighbour eigenvalue spacing (see \cite{sarika} and references therein).  The eigenvalues of the adjacency matrix are first ordered  as $\lambda_{1}\leq\lambda_{2}\leq\ldots\leq\lambda_{N}$, where N is the size of the network.  
Then the  ``unfolding''  procedure is performed, which can be accomplished using the transformations $\xi_{i}=N_{av}(\lambda)$, where $N_{av}(\lambda)$ is the cumulative distribution of the eigenvalues. For a discrete spectrum, the transformation enables to extract  the ``smooth'' part  from the sample spectrum and separate it the ``staircase'' in the cumulative distribution.
 The analytical form of the distribution $N_{av}$ is obtained using polynomial fit of ``smooth'' part. The spacing distribution $P(s)$ for the random-matrix models is then defined as the probability of finding the nearest-neighbor eigenvalue in the spectrum at the  distance $s$. For the unfolded eigenvalues  the distance is defined as  $s_{i}=\xi_{i+1}-\xi_{i}$. In the random matrix theory $P(s)$ is described by Brody distribution 
\begin{equation}
P=As^{\beta}e^{\alpha s^{\beta+1}} \ , \label{brody_form}
\end{equation}
where $A=(1+\beta)$ and $\alpha=[\Gamma(\frac{\beta+2}{\beta+1})]^{\beta+1}$. 
Brody distribution is an empirical formula characterized by one parameter $\beta$. For the case of Gaussian random matrices \cite{mehta} and various model networks \cite{sarika} it appears that  $\beta\approx1$. 
We perform the same procedure described above for four types of networks with six modules and with different clustering and average connectivity. The spacing distributions are shown in Fig.\ref{brody} compared to the analytic form of the Brody distribution in Eq.\ (\ref{brody_form}) with $\beta=1$. The results show that within the numerical error,  even though clustering, modularity and average connectivity influence the spectral density of the networks, the spacing  distribution remains unchanged and can be fitted with the same parameter $\beta \sim 1$.

\section{Conclusion} \label{sec:conclusion}

We have simulated high density traffic of information packets with local search and queuing at nodes \cite{tadic2007} on two types of networks with higher structures---modules of different sizes and scale-free internal structure. One of the goals was to emphasize the role of modular structure against the structure of the underlying network connecting these modules. Our results suggest that the network composed of several modules with clustered scale-free structure can bear much larger traffic density before jamming occurs, compared to the network of the same size but  with a single module of the  same structure. However, the traffic efficiency, measured with the statistical parameters and the scaling exponents, is reduced and strongly dependent on the way the modules are interconnected. Particularly, the slope of the travel time distribution is reflecting the structure of the connective network. Whereas, the traffic jamming first occurs in the modules due to the trapping of the packets within a module away from their destinations.  

Another aspect of this work concerns the reversed problem: recovering the network structure from the correlations between  the traffic time series. In our approach we generate the network-wide time series (traffic signals at each node) on pre-defined network structures, and have demonstrated how some standard filtering procedures works in the presence of modules. Particularly, the internal structure of the modules (presence of hubs and hierarchy between the nodes) induces spurious correlations which are elusive for the filtering methods.  Our results indicated that additional input about how the modules might be structured is necessary in order to increase the validity of the filtering methods.  
Two cases should be differentiated: One, when the network is fully partitioned into subgraphs (modules) of similar structure, like our Net269,  and the other, when the modules are immersed into connecting network which has a nontrivial structure by itself. 
Our findings are corroborated with the analysis of the eigenvalue spectra of Laplacian for both the original and the correlation (filtered) matrices. Specifically, we have considered the spectra of the normalized Laplacian of these matrices, which has the bounded spectrum in the range [0,2] and it is suitable for the comparison:  When the filtering is improved, the spectra of the filtered correlation matrix converge towards the spectra of the real connectivity matrix. The convergence  first occurs in the edges of the spectrum, and the eigenvectors related with the lowest nonzero eigenvalues localize on the modules, although the number of links between these modules can still be unrealistically large. We have shown that the  efficiency of the procedure is strongly dependent on internal homogeneity of these modules. More homogeneous modules can be detected with higher accuracy.  These results may have implications for the real complex systems, such as gene expressions \cite{collins2003,zivkovic2006,eshel2008}, or stock market data \cite{econophys}, where the time series are used to reconstruct the (unknown) underlying network structure. Our message is that, in contrast to local node connectivity, the mesoscopic structure (functional modules) can be identified with better accuracy.

{\bf Acknowledgments} Research supported in part by the program  P1-0044
(Slovenia) and  the national project OI141035 (Serbia), and the international projects BI-RS\-/\-08-09-047  and  MRTN-CT-2004-005728,  and  COST-P10 action.
The numerical results were obtained on the computer system of the Department of theoretical physics, Jo\v zef Stefan Institute, Ljubljana,  
and on the AEGIS
e-Infrastructure, supported in part by EU FP6 and FP7 projects CX-CMCS, EGEE-III and SEE-GRID-SCI,
at the Scientific Computic Laboratory, Institute of Physics, Belgrade.

\bibliographystyle{epj}
\bibliography{biblio_e.bib}
\end{document}